\documentclass[twocolumn,showpacs,preprintnumbers,amsmath,amssymb]{revtex4}
\usepackage{graphicx}% Include figure files
\usepackage{dcolumn}% Align table columns on decimal point
\usepackage{bm}% bold math
\newcommand{\half}{\frac{1}{2}}
\newcommand{\thalf}{\frac{3}{2}}
\newcommand{\inner}[2]{{\bm #1}\cdot {\bm #2}}
\newcommand{\del}{\partial}
\newcommand{\Slash}[1]{{/\hspace{-0.23cm}{#1}}}
\newcommand{\nn}{\nonumber}
\begin{document}

%\preprint{APS/123-QED}

\title{
%The $\Delta(1232)$ and its Chiral Partners in Quenched QCD
$\pi N$ and $\pi\pi N$ Couplings of the $\Delta(1232)$ and its Chiral Partners
%Linear Sigma Model with tNucleon,
%Interactions as a Consequence of their Interpolating Fields
}% Force line breaks with \\

\author{K. Nagata$^1$}\email{nagata@phys.cycu.edu.tw,nagata@rcnp.osaka-u.ac.jp}
\author{A. Hosaka$^2$}
\author{V. Dmitra\v sinovi\' c$^3$}
%\altaffiliation[Also at ]{}

\affiliation{
$^1$Department of Physics, Chung-Yuan Christian University,Chung-Li 320, Taiwan\\
$^2$Research Center for Nuclear Physics, Osaka University, Mihogaoka 10-1, Osaka 567-0047, Japan\\
$^3$Vin\v ca Institute of Nuclear Sciences, lab 010, 
P.O.Box 522, 11001 Beograd, Serbia
}%
% \affiliation{%
% $^2$Research Center for Nuclear Physics, Osaka University, Mihogaoka 10-1, Osaka 567-0047, Japan
% }%
% \affiliation{
% $^3$Vin\v ca Institute of Nuclear Sciences, lab 010, 
% P.O.Box 522, 11001 Beograd, Serbia
% }%

\date{\today}% It is always \today, today,

%==============
\begin{abstract}
We investigate the interactions and chiral properties of the four 
spin-$\thalf$-baryons : $N^-(D_{13})$, $N^+(P_{13})$,
$\Delta^+(P_{33})$ and $\Delta^-(D_{33})$ together with the nucleon.
We construct the $SU(2)_R \times SU(2)_L$ invariant interactions between the
spin-$\half$ and -$\thalf$ baryons with the aid of a new, specially
developed spin and isospin projection technique for these baryon
 fields, where the chiral invariant interactions contain 
one- and two-pion couplings.
We obtain simple relations for the coupling
constants of the one- and two-pion spin $\half-\thalf$ transitions terms. 
The relation for the one-pion interactions reasonably agrees with the experiments, which 
suggests that these spin-$\thalf$ baryons are chiral partners.
\end{abstract}
%==============
%\pacs{12.39.Fe, 14.20.Gk}
% 13.30.Eg: Hadronic decays of baryon, 14.65.Bt: light quark, 12.39.Fe:chiral Lagrangian
% 14.20.Gk: baryons

\maketitle                 % Produces the title.

Chiral symmetry $SU(2)_R \times SU(2)_L$ is a key property of the strong interaction. 
When the spontaneous breakdown $SU(2)_R\times SU(2)_L \to SU(2)_V$ occur, the broken symmetry
plays a dynamical role in various scattering processes involving its Nambu-Goldstone bosons,
i.e. the pions. 
Hadrons are then classified by isospin multiplets of the residual symmetry.
If chiral symmetry is restored at high temperature or density, hadrons
should form degenerate multiplets as classified by the full chiral group representations 
$(I_R, I_L)$, where $I_R(I_L)$ is isospin for the $SU(2)_R(SU(2)_L)$ group. 
Even in the broken world we may expect particles are expressed by one of chiral multiplets
or by their simple superpositions~\cite{Weinberg:1990xn}. A familiar example is for the chiral mesons $(\sigma,\vec{\pi})$ 
and for the vector mesons $(\vec{\rho},\vec{a}_1)$. 
With regards to the baryons, the role of chiral symmetry in the classification has
not been explored much, as one does not know which
hadrons would form which chiral multiplets as degenerate partners.

The linear realization of the chiral symmetry offers two merits.
First, hadrons in the same chiral multiplet but with different isospins are related 
by the larger symmetry $SU(2)_R\times SU(2)_L$ than 
$SU(2)_V$.  This can help reducing the number of the free parameters.
Second it is easy to investigate the property changes towards the 
chiral restoration as functions of the chiral condensate. 
Having these advantages, the purpose of this paper is to investigate 
the properties of baryons with respecting the chiral symmetry 
$SU(2)_R \times SU(2)_L$, especially we focus on baryon resonances.

It is particularly interesting that the recent studies~\cite{Leinweber:1992hy,Zanotti:2003fx,Leinweber:2004it,Sasaki:2005ug,Zhou:2006xe,Lee:2006bu} 
show that the $\Delta^+_{P_{33}}(1232)$ and $N^-_{D_{13}}(1520)$ resonances are 
qualitatively reproduced in the {\it quenched} lattice QCD, 
which validates to some extent the empirical assumption
that the baryons are dominated by their 3$q$ Fock components. 
Recently, we clarified the relation between the chiral multiplets 
and the quark structures~\cite{Nagata:2007di}.
For instance, interpolating fields used in Ref.~\cite{Leinweber:1992hy,Zanotti:2003fx,Leinweber:2004it,Sasaki:2005ug,Zhou:2006xe,Lee:2006bu} 
belong to the chiral multiplet $(1,\half)\oplus(\half,1)$~\cite{Nagata:2007di}. 
This is our starting assumption where a set of spin-$\thalf$ baryons form 
the chiral multiplet as chiral partners.
We extend this idea to include two other four-star resonances, the $N^+_{P_{13}}(1720)$ and
$\Delta^-_{D_{33}}(1700)$, following  Jido et. al.~\cite{Jido:1999hd}, 
where the four spin-$\thalf$ baryons form a certain set of chiral multiplets, so-called Quartet scheme.
Ref.~\cite{Jido:1999hd} mostly discussed the interactions between 
these chiral multiplets with the same spins, but not with the different spins especially 
the nucleon.
Inclusion of the nucleon enables us to testify such a framework in comparison with the 
experimental data for not only masses but also other quantities related to 
the dynamical processes such as resonance decays and scatterings.

In this Letter, we construct an effective Lagrangian for four types of four-star resonances,
$\Delta(1232)$, $N(1520)$ $N(1720)$ and $\Delta(1700)$ together 
with the nucleon and investigate the structures of the one- and two-pion coupling strengths.
We derive a relation among the one-pion coupling constants of the four baryon resonances, 
which agree well with the experimental data. We also consider the property changes of the one-pion 
couplings towards the chiral restoration.

We begin with the nucleon's chiral multiplet for which there are two
possibilities $(\half,0)\oplus(0,\half)$ and $(1,\half)\oplus
(\half,1)$ when the nucleon is a three-quark field.
We ssume the nucleon to be the fundamental representation.
There are also two possible chiral representations for the
$\Delta(1232)$: $(1,\half) \oplus (\half,1)$ and
$(\thalf,0)\oplus (0,\thalf)$.
We choose the former case as is commonly used to describe the
$\Delta(1232)$, as well as in Ref.~\cite{Jido:1999hd}.
Now we define two types of the diquarks;
 a Lorenz vector iso-scalar diquark $V^\mu$ ($I(J)^P=0(1)^-$)and  axial-vector 
iso-vector diquark $A^{\mu i}$ ($1(1)^+$),
\begin{subequations}
\begin{align}
V^{\mu } &= \tilde{q}  \gamma^\mu q, \\
A^{\mu i} &=\tilde{q}  \gamma^\mu\gamma_5 \tau^i q,
\end{align} 
\end{subequations}
where $\tilde{q}=q^T C (i\tau_2)\gamma_5$ is a transposed quark field.
These diquarks form the chiral multiplet $(\half,\half)$, just 
like the $\sigma$ and $\vec{\pi}$ mesons, which 
is a key ingredient in our construction of the chiral 
invariant interactions. We will come back to this point later.
There is one possible operator for $I(J)=\thalf(\thalf)$,
\begin{subequations} 
\begin{align}
\Delta_4^{\mu i}&=A_\nu^j \Gamma^{\mu\nu}_{3/2} P^{ij}_{3/2} q,
\end{align}
and two for $I(J)=\half(\thalf)$,
\begin{align}
N_V^{\mu}&= V_\nu \Gamma^{\mu\nu}_{3/2} \gamma_5 q,\\
N_A^{\mu}&= A_\nu^i \Gamma^{\mu\nu}_{3/2} \tau^i q.
\end{align}
\label{eq:4Mar08eq1}
\end{subequations}
Here the isospin-$\thalf$ $P^{ij}_{3/2}$ and the isospin-$\half$ projection operator
$P^{ij}_{1/2}$, satisfy $\delta^{ij} = P^{ij}_{3/2} + P^{ij}_{1/2}$~\cite{Nagata:2007di}.
Similarly the spin-$\thalf$ projection operator $\Gamma^{\mu\nu}_{3/2}$, with the
spin $\half$ projection operator $\Gamma^{\mu\nu}_{1/2}$, satisfy
the completeness relation $g^{\mu\nu}=\Gamma^{\mu\nu}_{3/2} +
\Gamma^{\mu\nu}_{1/2}$. Note that with the spin-$\thalf$ projection operator 
 the baryon fields still contain four fictitious spin-$\half$ components.
 However, the chiral transformation properties does not depend on the 
 choice of the spin projection operators, the local or non-local type
~\cite{Nagata:2007di}. 
Our strategy is  firstly use the local projection operators in 
the construction of the Lagrangian, later eliminate the spin-$\half$  components
in the calculations of the physical quantities, the one-pion decays in the 
present context.

Taking into account the normalization and Pauli-principle, which is 
implemented by the Fierz transformation, we define the baryon 
fields as
\begin{subequations}
\begin{align}
\Delta_1^{\mu i} &=\frac{\Delta_4^{\mu i}}{2},\\
N_1^\mu &=\frac{\sqrt{3}}{2} \frac{N_V^\mu}{2}+\half \frac{N_A^\mu}{2\sqrt{3}},
\end{align}
\end{subequations}
where we separate the coefficients to show explicitly the  
normalized baryon fields $\Delta_4^{\mu i}/2, N_V^\mu/2$ and $N_A^\mu /2\sqrt{3}$.
Note that the mixing between $N_V^\mu$ and $N_A^\mu$ results from the 
chiral transformations of $V^\mu$ and $A^{\mu i}$, and the mixing angle 
between $N_V^\mu$ and $N_A^\mu$ are determined by the Fierz 
transformation~\cite{Nagata:2007di}. 
The chiral transformation properties are straightforwardly given by 
\begin{subequations}
\begin{align}
\delta_5^{\vec{a}} N_1^\mu =& \frac53 i \inner{a}{\tau} \gamma_5
N_1^\mu + \frac{4}{\sqrt{3}} i \gamma_5 \inner{a}{\Delta_1^\mu},
\label{eq:chiralnmu}\\
\delta_5^{\vec{a}} \Delta_1^{\mu i} =& \frac{4}{\sqrt{3}} i
\gamma_5 a^j P^{ij}_{3/2} N_1^\mu
- \frac23 i \tau^i \gamma_5 \inner{a}{\Delta_1^\mu}\nn \\
+& i\inner{a}{\tau}\gamma_5\Delta_1^{\mu i},
\label{eq:chiraldelmu}
\end{align}%
\label{eq:su2a1}%
\end{subequations}%
implying that a set of $N_1^\mu$ and $\Delta_1^{\mu i}$ 
form the chiral multiplet $(1,\half)\oplus (\half,1)$.

Even after establishing these chiral transformations, it is a
non-trivial task to build chirally invariant interactions from
these fields, so we shall 
develop a new method to project out the good spin and isospin
parts from chiral invariant operators containing reducible products of
three-quark fields. 
This projection technique is performed in two
steps. Firstly we adopt a multiple product of meson 
and baryon operators described as direct products of quark bi-linears
(diquarks) and a quark. Here the equivalence of the chiral
properties of $(V^\mu, A^{\mu i})$ and $(\sigma, \pi^i)$ 
has been used. Since such composite operators are reducible
under both the chiral and the spin and isospin transformations,
we perform the decomposition into irreducible parts containing only spin and
isospin-projected baryons.

As an illustration let us consider the vector and axial-vector
diquarks $(V^\mu, A^{\mu i})$. As explained,  they belong to the chiral
multiplet $(\half,\half)$ similar to $(\sigma,\vec{\pi})$. 
Therefore the $V_\mu^2 + A_\mu^2$ combination is a chiral scalar, which
immediately leads to the chiral invariant term
$\bar{q} (V_\mu^2 + A_\mu^2)U_5 q$, where $U_5 = \sigma +
i\gamma_5 \inner{\tau}{\pi}$. The direct products of a quark and
diquarks $V^\mu q$ and $A^{\mu i} q$ contain several kinds of
baryons with $I(J)=\half(\half)$, $\half(\thalf)$,
$\thalf(\thalf)$~\footnote{ $I(J)=\thalf(\half)$
state is  forbidden by the Pauli-principle~\cite{Nagata:2007di} 
when it is described as a local three quark state.}. 
The decomposition into irreducible spin and
isospin parts is carried out by using the completeness relations
of both the spin and isospin projection operators. The
resulting interaction Lagrangian is given by
\begin{align}
{\cal L}_{\pi B B}^1&=  g_1\left( \bar{\Delta}_{1\mu}^i U_5 \Delta_1^{\mu i}
-\frac{3}{4}\bar{N}_{1\mu} U_5 N_1^\mu \right. \nn \\
&+\left.\frac{1}{12} \bar{N}_{1\mu} \tau^i U_5 \tau^i N_1^\mu
+\frac{\sqrt{3}}{6}\bar{N}_{1\mu} \tau^i U_5 \Delta_1^{\mu i} \right).
\label{eq:chiralint1}
\end{align}
Note that the relative weights for $N_1^\mu$ and $\Delta_1^{\mu i}$ are
unambiguously fixed by Eq.~(\ref{eq:su2a1}) without dependence 
on any free parameters~\footnote{ 
Eq.~(\ref{eq:chiralint1}) predicts a mass relation
$m(\Delta_1^{\mu i}) : m(N_1^\mu) = 2 : 1$. There are no candidates for this sort of $N^*$ and
$\Delta$ in the observed spectrum~\cite{Yao:2006px}.}.

Now, following Ref.~\cite{Jido:1999hd}, we introduce a new
set of spin-$\thalf$ baryons $(N_2^\mu, \Delta_2^{\mu i})$ that
have the $SU(2)_A$ transformation properties opposite in sign to
those of $(N_1^\mu, \Delta_1^{\mu i})$, so called
mirror baryons~\cite{Lee:1973,DeTar:1988kn,Jido:1999hd,Jido:2001nt}. They
may appear due to some more complicated structures such as non-local nature 
of three-quark states and multiquark components. Here we assume the
existence of the mirror baryons without considering their internal structures
 in detail. The diagonal interactions for the mirror
baryons are easily obtained:
\begin{align}
{\cal L}_{\pi B B}^2 &=  g_2\left( \bar{\Delta}_{2\mu}^i
U_5^\dagger \Delta_2^{\mu i}
-\frac{3}{4}\bar{N}_{2\mu} U_5^\dagger N_2^\mu \right. \nn \\
&+ \left.\frac{1}{12} \bar{N}_{2\mu} \tau^i U_5^\dagger \tau^i
N_2^\mu +\frac{\sqrt{3}}{6}\bar{N}_{2\mu} \tau^i U_5^\dagger
\Delta_2^{\mu i} \right). \label{eq:chiralint2}
\end{align}
In addition, the following mass terms are allowed
\begin{align}
{\cal L}_{BB} = - m_0 \left(\bar{\Delta}_{1 \mu }^i \Delta_2^{\mu
i} + \bar{N}_{1\mu} N_2^\mu \right).
\label{eq:chiralint3}
\end{align}
In contrast to Eqs.~(\ref{eq:chiralint1}) and (\ref{eq:chiralint2}), 
the mixings between $N_1^\mu (\Delta_1^{\mu i})$ and $N_2^\mu (\Delta_2^{\mu i})$
occur only after the mass diagonalization when the so-called mirror mass $m_0$ is
finite~\cite{DeTar:1988kn,Jido:1999hd,Jido:2001nt}. 
Combining Eqs.~(\ref{eq:chiralint1})$\sim$(\ref{eq:chiralint3}),
the quartet scheme of Jido et. al.~\cite{Jido:1999hd} is exactly
reproduced.

Next, we include the nucleon and its couplings with
the spin$-\thalf$ baryons, which is
new in this work. Similarly to the above discussion,
 $(V^\mu,\vec{A}^{\mu})$ and $(\sigma,\vec{\pi})$ form a chiral
scalar $\sigma V_\mu + i \inner{\pi}{A_\mu}$. Hence we find two
chirally invariant interactions: (1) $\bar{N}  U_5 [
(\del^\mu\sigma) V_\mu + i(\del^\mu\inner{\pi)}{A_\mu}]q$, and (2)
$\bar{N} (\del^\mu U_5) (\sigma V_\mu + i\inner{\pi}{A_\mu})q$.
With the irreducible decomposition, we obtain
\begin{align}
{\cal L}_{\pi N B}^1&= \frac{g_3}{\Lambda^2}
\left[\bar{N} U_5 (i\del_\mu \pi^i) \Delta^{\mu i}_1\right.
\nn \\
 &+ \left. \frac{\sqrt{3}}{2} \bar{N} U_5 ( \gamma_5\del_\mu \sigma
 + \frac{i}{3}\del_\mu \inner{\pi}{\tau})N^\mu_1\right],
\label{eq:chiralint4}\\
{\cal L}_{\pi N B}^2 &= \frac{g_4}{\Lambda^2} \left[\bar{N}
(\del_\mu U_5) (i  \pi^i)\Delta^{\mu i}_1\right.\nn \\
&+ \left.\frac{\sqrt{3}}{2} \bar{N} (\del_\mu U_5) (\gamma_5 \sigma
+\frac{i}{3} \inner{\pi}{\tau})N^\mu_1\right],
\label{eq:chiralint5}
\end{align}
where the dimensional parameter $\Lambda$ is introduced to
keep the coupling constants dimensionless. We neglect the
higher-order terms for the nucleon $\bar{N} U_5
\Slash{\del} U_5^\dagger N$ and $\bar{N}(\del^\mu U_5) U_5^\dagger
\gamma_\mu N$.
For the mirror baryons, we obtain the single-meson coupling
\begin{align}
{\cal L}_{\pi N B}^3 &= \frac{g_5}{\Lambda}
\left(\bar{N}(i\del^\mu \pi^i) \Delta_{2\mu }^i \right. \nn\\
&- \left. \frac{\sqrt{3}}{2}\bar{N} \del^\mu(\gamma_5\sigma-\frac{1}{3}
i \inner{\tau}{\pi})N_{2\mu} \right).
\label{eq:chiralint6}
\end{align}
Again, we neglect the nucleon term $\bar{N} \Slash{\del}
U_5^\dagger N_m$, where $N_m$ is another nucleon field having the
mirror properties.
Note that the interactions Eqs.~(\ref{eq:chiralint4}) 
and (\ref{eq:chiralint5}) involve two mesons, while Eq.~(\ref{eq:chiralint6})
contains only the single meson couplings.

%=============================================
\begin{table}[tbh]
\begin{center}
\caption{Masses (second column) and Coupling constants (third column).
For masses, we follow Jido et. 
al~\cite{Jido:1999hd}. The experimental values are taken from the PDG 
tables~\cite{Yao:2006px}. The experiments determine only the absolute 
values of the coupling constants, the positive values are our assumption.}
\begin{tabular}{ccccc}
\hline \hline
 States & Masses [MeV] &  $g_{\pi NB}/\Lambda$ [MeV$^{-1}$]& $\Gamma_{B\to \pi N}$ [MeV] \\
 & Theo (Exp)  & Theo (Exp)\\
\hline
$\Delta_{+}^{\mu i}$   ($P_{33}$) & 1320 (1232) & 15   (16) & 118\\
$\Delta_{-}^{\mu i}$   ($D_{33}$) & 1770 (1700) & 9.2 (9.5) &  45\\
$N_{-}^{\mu}$          ($D_{13}$) & 1430 (1520) & 9.4 (8.6) &  69\\
$N_{+}^{\mu}$          ($P_{13}$) & 1660 (1720) & 2.4 (2.4) &  30\\
\hline \hline
 & & $m_0=1550$ & $g_1=g_2=2.4$ \\
 & $g_5/\Lambda=17$ & $g_3 f_\pi /\Lambda^2 =4.2$ & $g_4 f_\pi /\Lambda^2 =8.2 $  \\
\hline\hline
\end{tabular}
\label{tab:input1}
\end{center}
\end{table}
%=============================================

Having constructed the Lagrangian with the nucleon and spin$-\thalf$ baryons,
let us determine the parameters $g_{1,2}$ and $m_0$, following Ref.~\cite{Jido:1999hd}. 
The results are shown in Table~\ref{tab:input1}~\footnote{In Fig.1 of Jido et. al. the state labeled
as 1770 and 1660 ought to be turned upside down.}.
After the diagonalization of the mass term and the corresponding parity (re)definition, the
mass eigen-states are obtained as: for the $\Delta$s,
$\Delta_{+}^{\mu i} = (\Delta_1^{\mu i} + \Delta_2^{\mu
i})/\sqrt{2}$, $\Delta_{-}^{\mu i}=\gamma_5 (-\Delta_1^{\mu
i}+\Delta_2^{\mu i})/\sqrt{2}$, and for the $N^*$s,
$N_{-}^{\mu} = \gamma_5 (-N_1^\mu + N_2^\mu)/\sqrt{2}$,
$N_{+}^{\mu} = (N_1^\mu + N_2^\mu)/\sqrt{2}$, where the subscripts
$\pm$ denote the parity ~\footnote{The slight difference from Ref.~\cite{Jido:1999hd}
 is caused by our definition that all the basis 
$\Delta_{1,2}$ and $N_{1,2}$ have positive parity.}.

After the spontaneous breaking $SU(2)_R\times SU(2)_L\to SU(2)_V$, the one-pion
interactions in Eqs.~(\ref{eq:chiralint4})-(\ref{eq:chiralint6}) are reduced to
\begin{subequations}
\begin{align}
{\cal L}_{\pi N B}&= \frac{g_{\pi N \Delta^+}}{\Lambda}\bar{N} (i\del_\mu \pi^i)
\Delta_+^{\mu i }+ \frac{g_{\pi N \Delta^-}}{\Lambda}\bar{N} (i \gamma_5
\del_\mu \pi^i)\Delta_-^{\mu i}\nn \\
&+ \frac{g_{\pi N N^{*-}}}{\Lambda}\bar{N}(i\gamma_5 \del_\mu
\inner{\pi}{\tau}) N_-^{\mu}\nn \\
&+ \frac{g_{\pi N N^{*+}}}{\Lambda} \bar{N}(i \del_\mu
\inner{\pi}{\tau}) N_+^{\mu},
\label{eq:chiralint7}
\end{align}
where the  coupling constants are given by 
\begin{align}
g_{\pi N \Delta^\pm} &=\frac{1}{\sqrt{2}\Lambda}(g_5\Lambda \pm g_3 f_\pi), \\
g_{\pi N N^{*\pm}} &=\frac{\sqrt{6}}{12\Lambda}\left(  (g_3 + 3g_4) f_\pi
\mp g_5\Lambda\right).
\end{align}
\label{eq:coupling}
\end{subequations}
The three coupling constants $g_{3,4,5}$ are determined from
the one-pion decay widths of the resonances as shown in Table~\ref{tab:input1}. We obtain quantitatively 
reasonable results for all the four coupling constants Eqs.~(\ref{eq:coupling}).
Eliminating $g_{3,4,5}$ from Eqs.~(\ref{eq:coupling}), we obtain 
a new relation:
\begin{align}
(g_{\pi N \Delta^+} + g_{\pi N \Delta^-})=2\sqrt{3}(g_{\pi N
N^{*-}} - g_{\pi N N^{*+}}),
\label{eq:couplingrel}
\end{align}
which satisfies the experimental data with a numerical error 
of less than $10 \%$. Considering the simplicity of the present 
description, this is an encouraging result suggesting that 
the spin$-\thalf$ baryons are the candidates of the chiral 
partners.

One of interesting properties of the present model is 
the two-pion contact terms, which are an inevitable
consequence of the chiral invariance. They involve only the
$g_3$ and $g_4$, while $g_5$, which is a leading contribution 
in the one-pion couplings, does not contribute to the two-pion 
couplings. The two-pion decay of $\Delta(1232)$ is therefore
suppressed by the smallness of the coupling constants as
compared to the one-pion decay. On top of this, the derivative
coupling causes an additional suppression of the two-pion decay
rate, due to the small final state pion momentum. Hence we can
expect strong suppression of the two-pion decay of $\Delta(1232)$.
Explicitly the two-pion contact interactions are given by
\begin{subequations}
\begin{align}
{\cal L}_{2\pi N B}&= \frac{1}{\sqrt{2}\Lambda^2}\bar{N} A_\mu^i
\Delta_{+}^{\mu i}-\frac{1}{\sqrt{2}\Lambda^2} \bar{N}
A_\mu^i \gamma_5 \Delta_{-}^{\mu i} \nn \\
&+  \frac{\sqrt{6}}{12\Lambda^2}\bar{N} B_\mu \gamma_5 N_{-}^{\mu} -
\frac{\sqrt{6}}{12\Lambda^2}\bar{N} B_\mu N_{+}^{\mu},
\end{align}
with
\begin{align}
A_\mu^i &= g_3(i\gamma_5\inner{\pi}{\tau})(i\del_\mu \pi^i)
+ g_4 (i\gamma_5 \del_\mu\inner{\pi}{\tau})(i\pi^i), \\
B_\mu &= g_3
(i\gamma_5\inner{\pi}{\tau})(i\del_\mu\inner{\pi}{\tau}) + g_4
(i\gamma_5\del_\mu \inner{\pi}{\tau})
(i\inner{\pi}{\tau}). \nn \\
\end{align}
\end{subequations}
Hence we obtain a relation 
between the 2-$\pi$ contact terms:
\begin{align}
|g_{2\pi N\Delta^+}| = |g_{2\pi N\Delta^-}|=2\sqrt{3}|g_{2\pi N
N^{*+}}| = 2\sqrt{3}|g_{2\pi N N^{*-}}|.
\end{align}
 In contrast to the $\Delta(1232)$ case,
it is expected that the two-pion contact term leads to 
larger contributions for other baryon resonances, because of the
larger final state pion momenta. In particular, the two-pion
coupling constants of $N^{*+} (\Delta^-)$
has the same magnitude as compared with that of $N^{*-}(\Delta^+)$, 
while the one-pion coupling constant is suppressed by the negative 
sign in Eqs.~(\ref{eq:coupling}). This qualitatively 
explains the observed feature of the two-pion decay enhancement in the decays of $N(1720)$
 and $\Delta(1700)$.
Due to the lack of other resonances, such as $\rho$-meson and $N(1440)$, we
do not consider this point in details.

%=======================================================
\begin{figure}
\includegraphics[width=6cm]{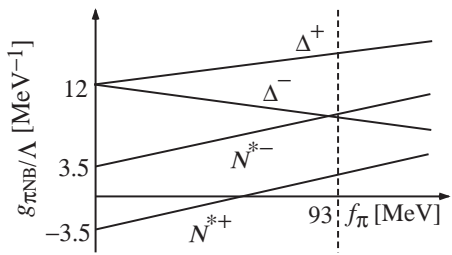}
\begin{minipage}{8cm}
\caption{The illustration of $f_{\pi}$ dependence of the one-pion 
coupling constants $g_{\pi N B}/\Lambda$ in the case (1).}
\label{fig:coupling}
\end{minipage}
\end{figure}
%=======================================================
As an application of our results to one of the recent interests, 
we consider a situation towards the  chiral restoration at high temperature 
or density.
We briefly consider the property changes of the one-pion coupling 
constants in two cases: (1) the scale parameter 
$\Lambda$ is constant and (2) $\Lambda =f_\pi$.  In the case (1),
as $f_\pi$ decreases, the three coupling constants $g_{\pi N
\Delta^+}$ and $g_{\pi N N^{*\pm}}$ decrease, while the remaining one
$g_{\pi N \Delta^-}$ increases. At $f_\pi=0$,
we obtain the simple relation $g_{\pi N \Delta^+} = g_{\pi N
\Delta^-} = \frac{\sqrt{3}}{6} g_{\pi N N^{*-}} =
-\frac{\sqrt{3}}{6} g_{\pi N N^{*+}}$, which is shown in Fig.~\ref{fig:coupling}. 
 In the case (2), all the coupling constants simply increase proportional to $f_\pi^{-1}$.
Eq.~(\ref{eq:couplingrel}) does not depend on the value of
$\Lambda$, hence it always holds in both cases.

In summary, we have investigated the chiral properties of 
four spin- $\thalf$ baryon  resonances together with the nucleon. 
We have constructed the effective interactions for the 
spin $\half-\thalf$  transition terms with the aid of
the spin and isospin projection formalism for the baryon
fields comprised of three quark fields. 
Of course, we can prove
the chiral invariance of the derived interactions directly from
the chiral transformation laws, but the results are general from
the group-theoretical point of view.
Within the $J=\thalf$ sector, the projection formalism reproduces the
Quartet scheme proposed by Jido et al.~\cite{Jido:1999hd}. In
addition, we derived the minimal chiral invariant one- and two-
 meson couplings with spin $\half-\thalf$ baryons. We
found that the one-pion couplings describing the spin
$\half-\thalf$ transitions are 
constrained by the chiral symmetry via the Eq.
(\ref{eq:couplingrel}), which quantitatively agrees with the
experiment. Considering the simplicity of our assumptions
on the effective Lagrangian, it is an remarkable result suggesting
that these baryons are chiral partners. In addition, we obtain 
chiral two-pion couplings, whose strengths are entirely determined by the
one-pion coupling constants. This enable us to predict two-pion 
decays of the resonances that can be tested in experiments. 
In this Letter we employed a new projection technique to
derive the effective chiral interaction Lagrangians between
baryons of different spin and isospin. 

We thank Prof. D. Jido for fruitful discussions. K.N and V.D thank 
Prof. H. Toki for hospitality in the stay at RCNP. 
K.N is supported by National Science Council (NSC) of Republic 
of China under grants No. NSC96-2119-M-002-001.

%\bibliography{ref_list}

\end{document}